\documentclass[conference,a4paper]{IEEEtran}

\usepackage{xcolor}
\usepackage{balance}

\ifCLASSINFOpdf
  \usepackage[pdftex]{graphicx}

\else

  \usepackage[dvips]{graphicx}

\fi

\ifCLASSOPTIONcompsoc
 \usepackage[caption=false,font=normalsize,labelfont=sf,textfont=sf]{subfig}
\else
 \usepackage[caption=false,font=footnotesize]{subfig}
\fi

\usepackage{multirow}
\usepackage{stfloats}
\usepackage{cite}
\usepackage{xcolor}
\usepackage{diagbox}
\usepackage{multirow}
\usepackage{multirow}
\usepackage{graphicx}
\usepackage{colortbl}
\usepackage{booktabs}
\usepackage{makecell}
\usepackage{float}
\usepackage{threeparttable}
\usepackage{graphicx}
\usepackage{bm}
\begin{document}

\title{\ A Ray-tracing and Deep Learning Fusion Super-resolution Modeling Method for Wireless Mobile Channel
}

\author{\IEEEauthorblockN{
Zhao Zhang\IEEEauthorrefmark{1}, 
Danping He\IEEEauthorrefmark{1}\IEEEauthorrefmark{2}, 
Xiping Wang\IEEEauthorrefmark{1},    
Ke Guan\IEEEauthorrefmark{1}\IEEEauthorrefmark{3}, 
Zhangdui Zhong\IEEEauthorrefmark{1}\IEEEauthorrefmark{4},
Jianwu Dou\IEEEauthorrefmark{5}\IEEEauthorrefmark{6}
}                                   
\IEEEauthorblockA{\IEEEauthorrefmark{1}
State Key Laboratory of Rail Traffic Control and Safety, Beijing Jiaotong University, 100044 Beijing, China}
\IEEEauthorblockA{\IEEEauthorrefmark{2}
Beijing Engineering Research Center of High-speed Railway Broadband Mobile Communications, 100044 Beijing, China}
\IEEEauthorblockA{\IEEEauthorrefmark{3}
Frontiers Science Center for Smart High-speed Railway System, 100044 Beijing, China}
\IEEEauthorblockA{\IEEEauthorrefmark{4}
Key Laboratory of Railway Industry of Broadband Mobile Information Communications, 100044 Beijing, China}
\IEEEauthorblockA{\IEEEauthorrefmark{5}
State Key Laboratory of Mobile Network and Mobile Multimedia Technology, 518055 Shenzhen, Guangdong, China}
\IEEEauthorblockA{\IEEEauthorrefmark{6}
ZTE Corporation, 518055 Shenzhen, Guangdong, China}

 \IEEEauthorblockA{\emph{*Corresponding author: Danping He, E-mail: hedanping@bjtu.edu.cn}}
}


\maketitle

\begin{abstract}
Mobile channel modeling has always been the core part for design, deployment and optimization of communication system, especially in 5G and beyond era. Deterministic channel modeling could precisely achieve mobile channel description, however with defects of equipment and time consuming. In this paper, we proposed a novel super resolution (SR) model for cluster characteristics prediction. The model is based on deep neural networks with residual connection. A series of simulations at 3.5 GHz are conducted by a three-dimensional ray tracing (RT) simulator in diverse scenarios. Cluster characteristics are extracted and corresponding data sets are constructed to train the model. Experiments demonstrate that the proposed SR approach could achieve better power and cluster location prediction performance than traditional interpolation method and the root mean square error (RMSE) drops by 51$\%$\ and 78$\%$ relatively.
Channel impulse response (CIR) is reconstructed based on cluster characteristics, which could match well with the multi-path component (MPC). The proposed method can be used to efficiently and accurately generate big data of mobile channel, which significantly reduces the computation time of RT-only.
\end{abstract}

\vskip0.5\baselineskip

\section{Introduction}
The global internet of vehicles market is projected to grow from $\$$ 95.62 billion in 2021 to $\$$ 369.60 billion in 2028 in forecast period \cite{c.elmohamed}. New communication services such as intra-vehicle, Vehicle-to-Everything (V2X) communication and Industrial Internet of Things (IIoT) have put forward more stringent requirements for mobile communication systems \cite{tan2021analysis}\cite{cheng2019cooperative}. 
It has been the consensus that the 5G and B5G will realize a high data-rate and ultra-reliable low-latency communication albeit with soaring demand. However, current 5G could hardly provide a guarantee in harsh electromagnetic and time-varying environments. To achieve B5G superior features for mobile communication in diverse scenarios, mobile channel modeling is imperatively needed to better understand channel characteristics, plan and optimize communication systems.

Channel modeling is the process of modeling the signal propagation mechanism to obtain an accurate channel description. Generally, geometry-based stochastic modeling (GBSM) and ray-tracing (RT) based deterministic modeling are two main modeling approaches for mobile channel. GBSM can theoretically generate channel impulse responses (CIRs) through assumed scattering geometry to analyze performance, but practically the CIRs are evaluated numerically via measurement and calculation. On the other hand, RT can generate highly accurate channel characteristics for specific scenario with the defect of computational complexity \cite{Qingtao2021high,Youping2021SBR,linxue2018Calibration}. Thereby, a fast and reliable channel characteristic generation method based on coarse-grained RT is desperately needed to accelerate the modeling process.

Mobile channel modeling not only refers to time-varying and non-stationary channels but also involves the influence of variant multi-path components and small-scale fading, which is more dominant. In this regard, cluster-based approaches are increasingly adopted in recent research especially since GBSM was proposed. In \cite{zhu2020ray}, a clustering and tracking algorithm was proposed and analyzed for vehicle-to-infrastructure channel. To recognize and track the clusters in time-varying channels, a clustering and tracking algorithm based on power-angle-spectrum is proposed and investigated \cite{huang2018power}. In \cite{zhao2022semi}, a semi-deterministic channel modeling method is presented based on RT and cluster. Many classical radio channel models are also based on the concept of cluster, e.g. COST 2100, 3GPP Spatial Channel Model and WINNER.

Deep learning (DL) based channel modeling methods are getting popular in recent years because of its excellent information integration and inferring ability. Numerous research including our previous work \cite{GLOBECOM2022} are dedicated on large scale channel characteristics prediction, e.g. pathloss, delay spread and number of clusters. Multi-layer perceptron artificial neural network is presented for path loss prediction in \cite{lina2021path}. The author in \cite{bai2018predicting} proposes a procedure of predicting channel characteristics based on convolutional neural network (CNN) for multi-dimensional millimeter wave channel characteristics prediction. However, small scale channel characteristics (e.g. power, angle and location of each cluster) related research based on DL are inadequate.

\begin{figure}
\setlength{\abovedisplayskip}{3pt}
\setlength{\belowdisplayskip}{3pt}
\centering
\includegraphics[width=0.9\columnwidth]{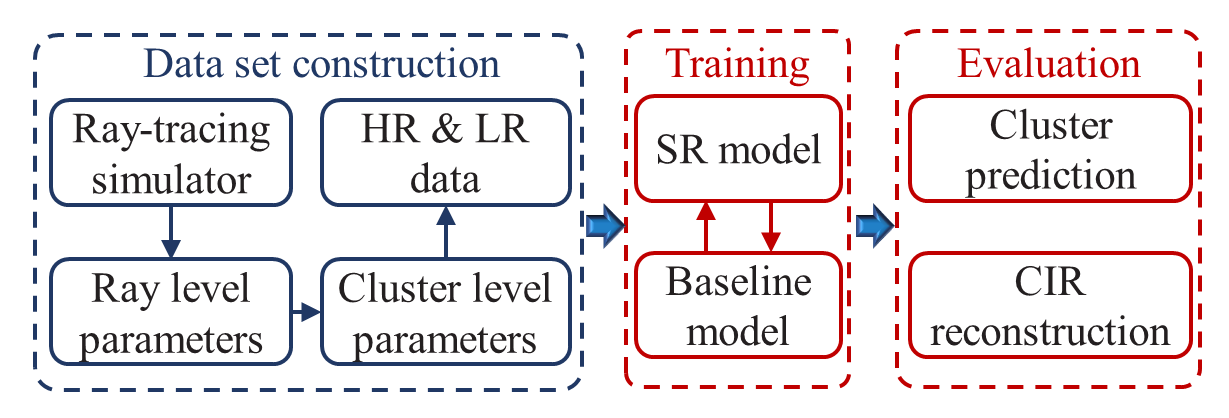}
\caption{Framework of the study.}
\label{fig:1}
\vspace{-0.2cm}
\end{figure}

In this paper, we propose a deep learning and ray tracing fusion super-resolution (SR) method for cluster characteristics, the power and location of each cluster more specifically. Overview of our work is shown in Fig. \ref{fig:1}. We first use ray-tracing simulator to generate ray level parameters, from which cluster level parameters are extracted utilizing the proposed clustering method. Subsequently, high resolution data and low resolution data are divided to train the deep learning SR model. Evaluation and comparison are implemented in cluster prediction and CIR reconstruction aspects. Specifically, we make the following contributions:
\begin{itemize}
 \item Massive RT simulation by self develop CloudRT \cite{he2018design} was conducted in four restored scenarios via SketchUp and 3D electronic map and multi-dimensional ray level characteristics data are generated.
 \item A novel object-based clustering method is presented for better rays division. Special filtering, tracking and segmenting method are utilized to extract crucial characteristics, unify cluster dataset and maintain continuity.
 \item Multi-layer deep learning model (MLL) is proposed to predict cluster characteristics more accurately. Residual connection and pre-upsampling techniques are integrated for eliminating the vanishing gradients problem and achieving better performance. Ablation study and generalization test demonstrated the necessity of adopted crucial techniques and adaptability of proposed model.
\end{itemize}

The remainder of this paper is organized as follows. Section II introduces the simulation configuration and data construction method. The proposed SR and baseline model are explained in Section III. Experiments and evaluations are implemented in section IV. Finally, conclusions are drawn in Section V.

\section{Simulation and data construction}
\subsection{Scenario Modeling}
\begin{figure}
\setlength{\abovedisplayskip}{3pt}
\setlength{\belowdisplayskip}{3pt}
\centering
\includegraphics[width=0.9\columnwidth]{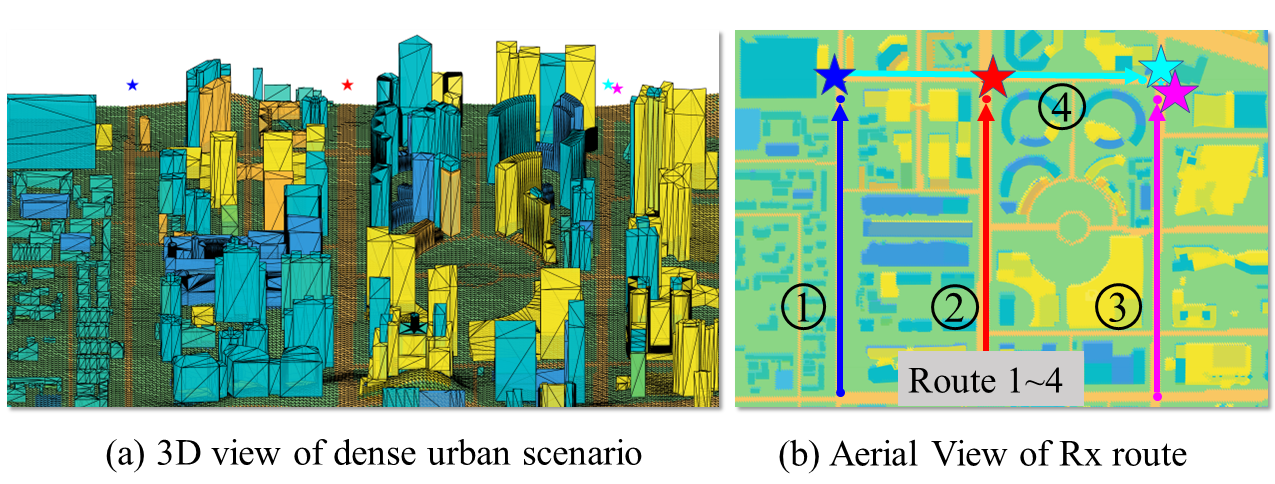}
\caption{Simulations in dense urban scenario.}
\label{fig:2}
\vspace{-0.2cm}
\end{figure}

\begin{figure}
\vspace{-0.2cm}
\setlength{\abovedisplayskip}{3pt}
\setlength{\belowdisplayskip}{3pt}
\centering
\includegraphics[width=0.9\columnwidth]{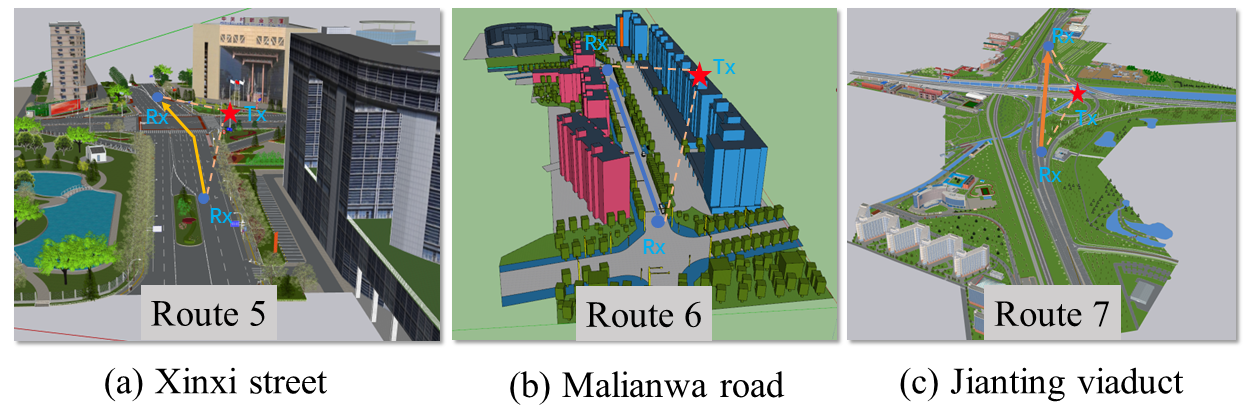}
\caption{Simulations in street scenarios.}
\label{fig:3}
\vspace{-0.2cm}
\end{figure}

The overall simulation is based on two kinds of scenarios. As shown in Fig. \ref{fig:2} and Fig. \ref{fig:3}, a dense urban scenario is based on 3D electronic map and three street scenarios are manually modeled via SketchUp. Note that all four scenarios are completely sourced from real environment, which are Central business district, Malianwa street and Jianting viaduct and Xinxi road in Beijing, respectively. The dense urban scenario mainly consists of different types of buildings and terrain (e.g. regular buildings, parallel buildings, dry land, green land). The street scenario also contains many fine scatterers, covering from cars to pedestrians and trees. 
As depicted, the transmitter (Tx) and route of receivers (Rx) are presented clearly. A total of 7 Rx routes in different color are simulated separately. It is noteworthy that Tx location is elaborately sited twice so that both LOS and NLOS scenes are simulated for route $1\sim4$. Only LOS scene is simulated for route $5\sim7$. Through simulation in various and close-to-real scenarios, the diversity of data is guaranteed in the case of limited data.

\subsection{Simulation Settings}
Based on aforementioned scenario and electromagnetic (EM) parameters provided by ITU-R P.1238-7, Self developed CloudRT platform is utilized to obtain channel characteristics data. 
Rx is located 2 meters above ground along the road at 1 m intervals. Tx is placed on the side of road, near the end of receiver's trajectory. The propagation mechanisms includes light of sight, scattering, reflection, penetration and diffraction. The detailed simulation configuration is shown in TABLE \ref{table1}. 

\begin{table}[!t]
\vspace{-0.2cm}
\setlength{\abovedisplayskip}{3pt}
\setlength{\belowdisplayskip}{3pt}
\renewcommand{\arraystretch}{1.3}
\caption{Simulation configuration}
\label{table1}
\centering
\scalebox{0.87}{
\begin{tabular}{|c||c|}
\hline
Parameter & Value \\
\hline
Carrier frequency [GHz] & 3.55\\
\hline
System bandwidth [MHz] & 100 \\
\hline
Tx transmit power [dBm] & 0.1 \\
\hline
Tx location & 5-10 m above ground \\
\hline
Rx location & 2 m above ground\\
\hline
Tx \& Rx attenna & Omni-directional vertical polarization \\
\hline
\end{tabular}}
\end{table}

\subsection{Clustering}

\begin{figure}[!t]
\setlength{\abovedisplayskip}{3pt}
\setlength{\belowdisplayskip}{3pt}
\centering
\includegraphics[width=1.0\columnwidth]{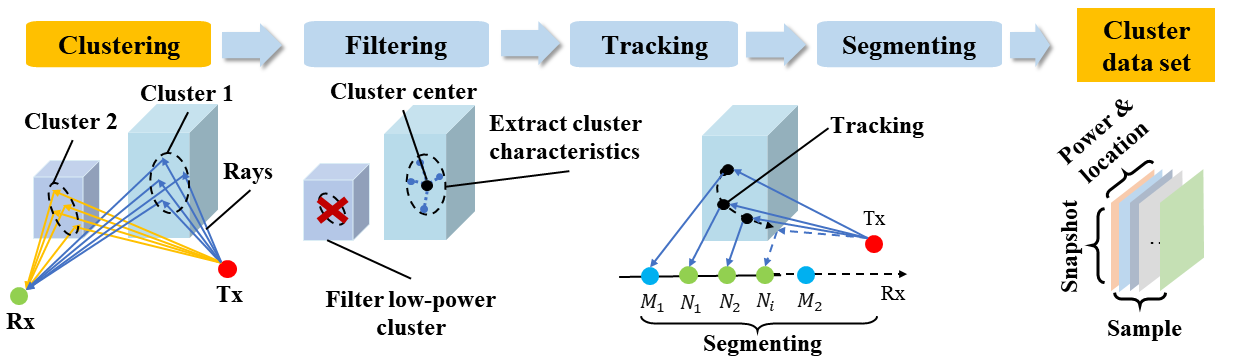}
\caption{Workflow of data process.}
\label{fig:4}
\end{figure}
\vspace{-0.1cm}
For a single Tx-Rx link, hundreds to thousands of rays can be traced, most of which share the similar delay, angle and power. Therefore, it is an accepted practice to equate similar rays as clusters and study channel characteristics on cluster level. 
The overall data process workflow is shown in Fig. \ref{fig:4}. Clustering is the process of cluster data generation based on rays. There are enormous approaches for clustering (e.g. power-angle-spectrum based clustering, power-delay profile based clustering). Based on RT, the propagation of each ray can be accurately acquired, including 3D coordinates, object and microfacet identity of reflection and scatter points with scene. 
Microfacet-based and object-based clustering method are hence considered and compared. In this work, rays that hit on the same object are grouped into a cluster, after which cluster characteristics including power and center locations are extracted for each snapshot. The author in \cite{huang2018power} proposed an algorithm for identifying and tracking multipath clusters, which also introduces the conception of power-weighted cluster centers, intra-cluster angle expansion and cluster shape. Similarly, we adopt the power-weighted cluster center and coherent superposition cluster power approach for filtering. The equation is as follows:

\begin{equation}\label{equ:1}
C(j) = \frac{\sum_{k=1}^K C(r_{k})P(r_{k})}{\sum_{k=1}^K P(r_{k})},r_{k}\in j 
\vspace{-0.2cm}
\end{equation}

\begin{equation}\label{equ:2}
\setlength{\abovedisplayskip}{1pt}
\setlength{\belowdisplayskip}{1pt}
P(j) = \sum_{k=1}^K P(r_{k}),r_{k}\in j 
\end{equation}

$C$, $P$ denotes the 3D coordinate of reflection or scatter point and power. $j$, $r_{k}$ denotes $j$th cluster and $k$th ray in this cluster.
Due to different cluster between snapshots, cluster tracking is implemented between snapshots for continuity. Every 17 adjacent snapshots are then segmented into a sample, exactly enough to implement SR test at scale 16. Within each sample, the number of clusters is compromised to a certain value to form a regular data structure. Data structure is shown in Fig. \ref{fig:4}, the data we refer here is the cluster power and intersections with scene. 
In total, 2024 samples were generated and combined to construct the dataset. After that, the input data are processed by down-sampling by certain SR scale factors.

\section{methodology}
\subsection{Problem Definition}
\begin{figure}
\setlength{\abovedisplayskip}{3pt}
\setlength{\belowdisplayskip}{3pt}
\centering
\includegraphics[width=0.9\columnwidth]{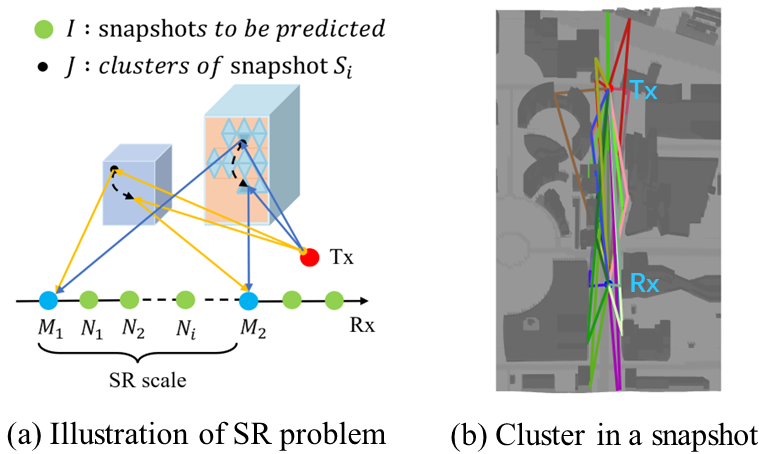}
\caption{SR problem and clustering result.}
\label{fig:5}
\vspace{-0.4cm}
\end{figure}

Super resolution means generate high resolution (HR) data $\hat{I}^{HR}$ from low resolution (LR) data $I^{LR}$. The objective of SR is to minimize the gap between $\hat{I}^{HR}$ and ground truth $I^{HR}$ while obtaining the best model parameters $\theta$, which is shown in (\ref{equ:3}) and (\ref{equ:4}) respectively.

\begin{equation}\label{equ:3}
\hat{I}^{HR} = \mathcal{F}(I^{LR}, \theta) 
\vspace{-0.2cm}
\end{equation}

\begin{equation}\label{equ:4}
\vspace{-0.2cm}
\setlength{\abovedisplayskip}{3pt}
\setlength{\belowdisplayskip}{3pt}
\theta = \mathop{\arg\min}\limits_{\theta}\mathcal{L}(I^{HR},\hat{I}^{HR})
\end{equation}
As shown in Fig. \ref{fig:5}(a), given a certain Rx track and SR scale factor $\delta$, M, N denote the known LR snapshots and unknown HR snapshots of receiver. $\delta-1$ snapshots between two adjacent M are to be predicted. There could be dozens of clusters for a single snapshot, as in Fig. \ref{fig:5}(b).  
The SR model intents to capture and predict the variation of cluster characteristics (e.g. the movement of cluster center present in black dotted line).

\subsection{Baseline model}
The baseline model is linear interpolation. Considering Rx track can be seen as straight line in small ranges, the linear interpolation method is as follows:

\begin{equation}\label{equ:5}
\setlength{\abovedisplayskip}{3pt}
\setlength{\belowdisplayskip}{3pt}
\renewcommand{\arraystretch}{2.1}
\begin{array}{c}
f_{LI}(N_{i,j})=\frac{||M_1N_i||}{||M_1M_2||}f(M_{2,j})+\frac{||N_iM_2||}{||M_1M_2||}f(M_{1,j}) \\ 
 \forall  i,j\in I,J
\end{array}
\end{equation}

where, $I$,$J$ describe the snapshots to be predicted and cluster of snapshot $S_i (S=M,N)$. $f$ represents cluster characteristics which can be 3D coordinate and power of cluster center, as in (\ref{equ:1}) and (\ref{equ:2}).

\subsection{Residual Network based Multi-layer Learning Model}
\begin{figure}
\centering
\includegraphics[width=0.9\columnwidth]{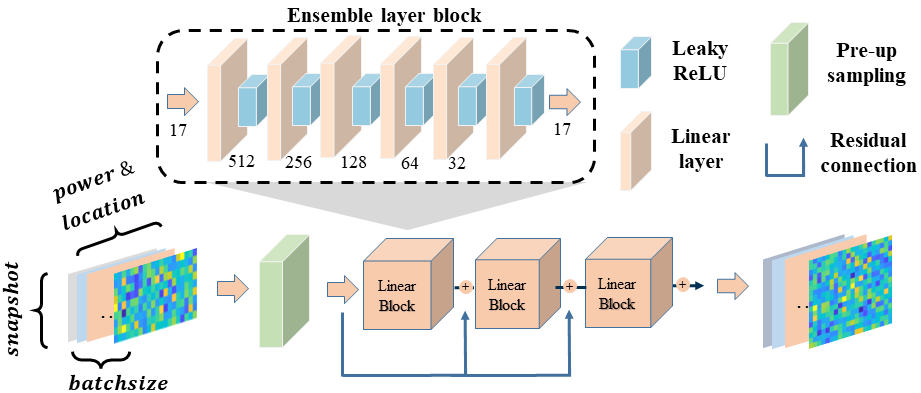}
\caption{The overview of proposed residual network based multi-layer deep learning model.}
\label{fig:6}
\end{figure}

Deep neural networks have achieved great success and high-quality reconstruction for image super-resolution. So the network needs to be designed very deep for a better mapping and inference between LR and HR data.
As illustrated in Fig. \ref{fig:6}, the multi-layer learning model is composed of three ensemble linear blocks (ELB). 
Six hidden layers with suddenly increasing and gradually declining dimension changes are designed in each ELB, thus iterative up-and-down change in feature dimension is forming to filter irrelevant information in input data. However, the vanishing gradients issue will become more apparent as the model deepens. Two techniques are utilized to eliminate this problem. First, residual connection, which has exhibited superior performance in computer vision problems, is added between each ELB to ease training process and accelerate convergence. Second, a pre-upsampling operation using baseline method is implemented, which outperforms transposed convolution layer demonstrated by previous experiment. Specifically, let $E$ denote the transform in ELB, the final cluster characteristics $f_{MLL}$ could be written as in (\ref{equ:6}).
The ELB number is set to be 3 for a balance of performance and training complexity. In addition, proposed model framework can be regarded as a general channel characteristics generation architecture that also performs well in our previous work \cite{GLOBECOM2022}.

\begin{equation}\label{equ:6}
\resizebox{1.00\hsize}{!}{$f_{MLL}(N_{i,j})=E^3(f_{LI}(N_{i,j}))+E^2(f_{LI}(N_{i,j}))+E(f_{LI}(N_{i,j}))$}
\end{equation}

\subsection{Loss Functions and Evaluation Metrics}
For SR task, only predicted snapshots ($N_i$) need to be evaluated for characteristics differences so the loss function $\mathcal{L}_{MLL}$ in training could be written as follows:

\begin{equation}\label{equ:7}
\mathcal{L}_{MLL}(\hat{I}^{HR},I^{HR})= \sum_{i=0}^I\sum_{j=0}^J\sum_{f}^{C,P}(\dot{f}_{MLL}(N_{i,j})-\dot{f}(N_{i,j}))^2
\end{equation}

According to previous experiment, L2 loss reduces characteristics error to lower level compared with L1 Loss. The prediction errors for different characteristics of each cluster in each snapshot will be added up successively for back propagation and parameter update.
Instead of birth and death prediction, we intend to train the model to better understand the evolution of clusters and correlation between snapshots. However, cluster birth and death occasionally arise among consecutive snapshots. Therefore, $f$ is transformed to $\dot{f}$ by multiplying the weighted matrix with value ${10^{-2}}$ for these inconsecutive clusters and 1.0 for normal clusters to achieve better training and evaluation. Weight values are evaluated from ${10^{-1}}$ to ${10^{-5}}$, and ${10^{-2}}$ is the optimal.
Absolute mean error (AME), mean absolute error (MAE) and root mean square error (RMSE) are basic evaluation metrics in this work.

\section{Experiment and evaluation}
\subsection{Training and Implementation Details}
In this study, training, validation and test experiments are conducted by PyTorch 1.9.0 on a core server with 1 NVIDIA RTX 3090 GPU, Intel Core i9-9900K CPU and 32 GB DDR4 RAM. To be noted, we elaborately divide the overall data as training, validation and test parts. The ratio of training set to validation set plus test set is about 5:1. Specifically, simulation results in dense urban scenarios route $1\sim4$ are divided into training set and validation set. Results in route $5\sim7$ are test set for generalization test.
The model is trained for 80 epochs before validation and test. The learning rate is set as $10^{-5}$. Adam optimizer is used for gradient descent. Experiments were carried out at SR scale factor 2, 4, 8 and 16.

\subsection{Performance of Proposed Model}
The best prediction results achieved by proposed model are illustrated in TABLE \ref{table3}, which exhibits the AME and RMSE of cluster power and location of cluster center. The training and validation of model are conducted in dense urban LOS and NLOS scenarios respectively. Prediction error is generally larger in LOS due to large quantity and severe variation of cluster. The evaluation metrics, AME and RMSE, are far smaller than baseline model, with error drops by 49$\sim$94$\%$ in LOS scene. It is noted worthy that the power prediction performance of proposed model deteriorates slightly in less harsh NLOS environment. We will further investigate this part in future research.

\begin{table}
\newcommand{\tabincell}[2]{\begin{tabular}{@{}#1@{}}#2\end{tabular}}
\renewcommand{\arraystretch}{1.2}
\caption{Super resolution performance of baseline and proposed model}
\label{table3}
\centering
\scalebox{0.9}{
\begin{tabular}{c||c|c|c|c|c}
\hline
\multicolumn{6}{c}{Absolute mean error (AME)} \\
\hline
 & & \multicolumn{2}{|c|}{LOS} & \multicolumn{2}{|c}{NLOS} \\
\hline
\tabincell{c}{scale} & \tabincell{c}{method} & \tabincell{c}{AME of \\ power \\ $[\rm dB]$} & \tabincell{c}{AME of \\location\\ $[\rm m]$} & \tabincell{c}{AME of \\ power\\ $[\rm dB]$}  & \tabincell{c}{AME of \\ location\\ $[\rm m]$} \\
\hline
\multirow{2}*{2} & Baseline & 1.57 & 2.09 & 0.73 & 1.52 \\
\cline{2-6}
~ & {\cellcolor[gray]{.85}}Proposed & {\cellcolor[gray]{.85}}0.80 & {\cellcolor[gray]{.85}}0.12 & {\cellcolor[gray]{.85}}0.87 & {\cellcolor[gray]{.85}}0.44 \\
\hline
\multirow{2}*{4} & Baseline & 1.66 & 2.18 & 0.81 & 1.65\\
\cline{2-6}
~ & {\cellcolor[gray]{.85}}Proposed & {\cellcolor[gray]{.85}}0.14 & {\cellcolor[gray]{.85}}0.25 & {\cellcolor[gray]{.85}}0.16 & {\cellcolor[gray]{.85}}0.45 \\
\hline
\multirow{2}*{8} & Baseline & 1.90 & 2.49 & 0.87 & 1.80 \\
\cline{2-6}
~ & {\cellcolor[gray]{.85}}Proposed & {\cellcolor[gray]{.85}}0.56 & {\cellcolor[gray]{.85}}0.31 & {\cellcolor[gray]{.85}}0.60 & {\cellcolor[gray]{.85}}0.30 \\
\hline
\multirow{2}*{16} & Baseline & 2.13 & 2.74 & 0.96 & 1.84\\
\cline{2-6}
~ & {\cellcolor[gray]{.85}}Proposed & {\cellcolor[gray]{.85}}0.08 & {\cellcolor[gray]{.85}}0.93 & {\cellcolor[gray]{.85}}0.22 & {\cellcolor[gray]{.85}}0.82 \\
\hline
\multicolumn{6}{c}{Root mean squared error (RMSE)} \\
\hline
 & & \multicolumn{2}{|c|}{LOS} & \multicolumn{2}{|c}{NLOS} \\
\hline
\tabincell{c}{scale} & \tabincell{c}{method} & \tabincell{c}{RMSE of \\ power \\ $[\rm dB]$} & \tabincell{c}{RMSE of \\location\\ $[\rm m]$} & \tabincell{c}{RMSE of \\ power\\ $[\rm dB]$}  & \tabincell{c}{RMSE of \\ location\\ $[\rm m]$} \\
\hline
\multirow{2}*{2} & Baseline & 9.71 & 13.20 & 6.99 & 8.83 \\
\cline{2-6}
~ & {\cellcolor[gray]{.85}}Proposed & {\cellcolor[gray]{.85}}4.81 & {\cellcolor[gray]{.85}}2.33 & {\cellcolor[gray]{.85}}4.63 & {\cellcolor[gray]{.85}}2.06 \\
\hline
\multirow{2}*{4} & Baseline & 10.32 & 13.25 & 7.63 & 8.24\\
\cline{2-6}
~ & {\cellcolor[gray]{.85}}Proposed & {\cellcolor[gray]{.85}}5.04 & {\cellcolor[gray]{.85}}2.56 & {\cellcolor[gray]{.85}}4.89 & {\cellcolor[gray]{.85}}2.19 \\
\hline
\multirow{2}*{8} & Baseline & 11.08 & 13.63 & 7.85 & 7.52 \\
\cline{2-6}
~ & {\cellcolor[gray]{.85}}Proposed & {\cellcolor[gray]{.85}}5.37 & {\cellcolor[gray]{.85}}2.92 & {\cellcolor[gray]{.85}}5.23 & {\cellcolor[gray]{.85}}2.25 \\
\hline
\multirow{2}*{16} & Baseline & 11.79 & 14.03 & 8.26 & 7.31\\
\cline{2-6}
~ & {\cellcolor[gray]{.85}}Proposed & {\cellcolor[gray]{.85}}5.70 & {\cellcolor[gray]{.85}}3.82 & {\cellcolor[gray]{.85}}5.51 & {\cellcolor[gray]{.85}}4.07 \\
\hline
\end{tabular}}
\end{table}


\subsection{Channel Impulse Response Reconstruction}
\begin{figure}
\centering
\includegraphics[width=0.85\columnwidth]{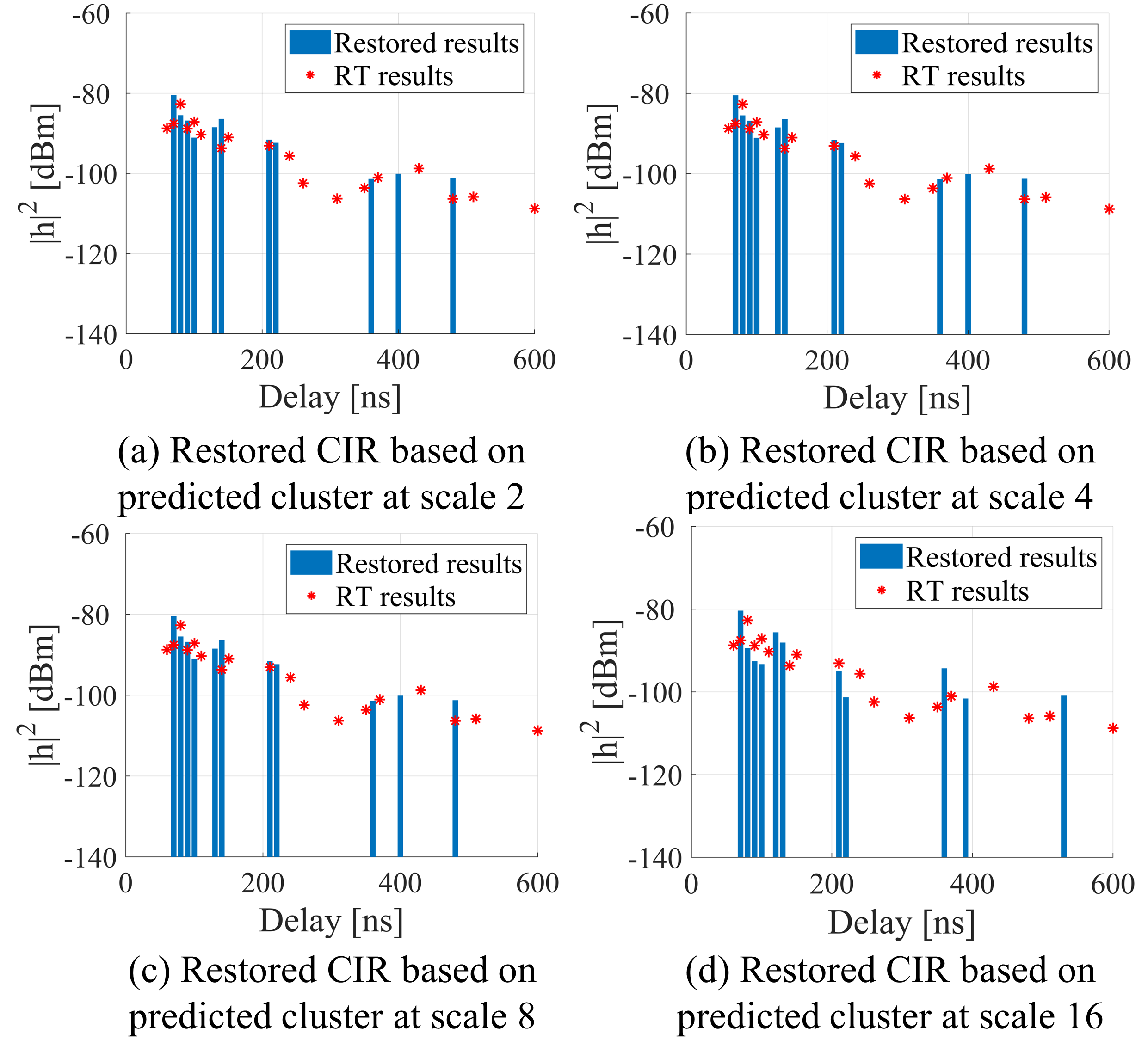}
\caption{Restored CIR at different scales.}
\label{fig:9}
\end{figure}

To better evaluate the SR performance for cluster characteristics, we regenerate the CIR based on predicted 3D positions and power of clusters. The simulated CIR indicated by the red asterisk is generated directly by RT. As can be seen in Fig. \ref{fig:9}, restored CIR could match most MPCs at different scales. Different from interpolation method, MLL model generates precise cluster characteristics that could restore CIR consistent with simulated at larger SR scales.

\subsection{Ablation Study and Generalization Test}
Ablation study was implemented to investigate the effectiveness of specific parts and designs in proposed model. Multiple linear layers with different hidden dimensions are integrated in ELB. By gradually increasing the layer number and its hidden dimension, the model could extract and learn a deeper variation of clusters. As demonstrated in TABLE \ref{table5}, the max hidden dimension in ELB is set to 512, which obtains the best performance. Residual connection is also indispensable to speed up convergence process and reduce errors, achieving more than 10$\%$ performance improvement.
Generalization test is also implemented at Jianting viaduct, Xinxi street and Malianwa road, as in TABLE \ref{table6}. Compared with the LOS results in dense urban scenarios, result is better in first two scenes and worse in Malianwa road. Without apparent model overfitting, it can be applied to other scenarios for channel modeling.

\begin{table}
\newcommand{\tabincell}[2]{\begin{tabular}{@{}#1@{}}#2\end{tabular}}
\renewcommand{\arraystretch}{1.16}
\caption{Cumulative super resolution error decline of cluster power}
\label{table5}
\centering

\setlength{\tabcolsep}{1.65mm}{
\scalebox{0.85}{
\begin{tabular}{c||c|c|c|c|c|c}
\hline
\multicolumn{7}{c}{ELB}\\
\hline
\tabincell{c}{Max hidden\\ dimension} & \makebox[0.025\textwidth][c]{32} & \makebox[0.015\textwidth][c]{64} & \makebox[0.015\textwidth][c]{128} & \makebox[0.015\textwidth][c]{256} & 512\textbf{ (ours)} &
\makebox[0.015\textwidth][c]{1024}\\
\hline
MAE & 0 & +1.9$\%$ & -9.1$\%$ & -20.9$\%$ & -26.4$\%$ & -24.3$\%$  \\
\hline
\multicolumn{7}{c}{RES} \\
\hline
Residual connection & \multicolumn{3}{|c|}{w/o} & \multicolumn{3}{|c}{w}\\
\hline
MAE & \multicolumn{3}{|c|}{0} & \multicolumn{3}{|c}{-11.9$\%$}\\
\hline
\end{tabular}}}
\vspace{-0.2cm}
\end{table}

\begin{table}[!t]
\newcommand{\tabincell}[2]{\begin{tabular}{@{}#1@{}}#2\end{tabular}}
\vspace{-0.2cm}
\renewcommand{\arraystretch}{1.16}
\caption{Super resolution performance (RMSE) in generalization test}
\label{table6}
\centering
\scalebox{0.85}{
\begin{tabular}{c|c|c|c|c|c|c}
\hline
 & \multicolumn{2}{|c|}{\makebox[0.02\textwidth][c]{Jianting viaduct}} & \multicolumn{2}{|c}{Xinxi street} & \multicolumn{2}{|c}{Malianwa road}\\
\hline
\tabincell{c}{scale} & \tabincell{c}{RMSE of \\ power\\ $[\rm dB]$} & \tabincell{c}{RMSE of \\location\\ $[\rm m]$} & \tabincell{c}{RMSE of \\ power\\ $[\rm dB]$} &  \tabincell{c}{RMSE of \\location\\ $[\rm m]$} & \tabincell{c}{RMSE of \\ power\\ $[\rm dB]$} & \tabincell{c}{RMSE of \\location\\ $[\rm m]$} \\
\hline

2 &  {\cellcolor[gray]{.85}}3.76 & {\cellcolor[gray]{.85}}2.63 & {\cellcolor[gray]{.85}}3.49 & {\cellcolor[gray]{.85}}2.70 & {\cellcolor[gray]{.85}}5.40 & {\cellcolor[gray]{.85}}5.59 \\
\hline

4  & {\cellcolor[gray]{.85}}3.97 & {\cellcolor[gray]{.85}}2.82 & {\cellcolor[gray]{.85}}3.94 & {\cellcolor[gray]{.85}}2.53 & {\cellcolor[gray]{.85}}6.19 &{\cellcolor[gray]{.85}}5.41\\
\hline

8  & {\cellcolor[gray]{.85}}4.29 & {\cellcolor[gray]{.85}}3.94 & {\cellcolor[gray]{.85}}3.88 & {\cellcolor[gray]{.85}}3.94 &{\cellcolor[gray]{.85}}5.93 & {\cellcolor[gray]{.85}}5.30\\
\hline
\end{tabular}}
\end{table}

\section{Conclusion}
In this paper, an efficient SR approach for cluster characteristics based on ray tracing and deep learning is proposed. Object-based clustering method is conducted to generate cluster characteristics. A multi-layer deep learning model is then proposed for cluster characteristic prediction. Based on LR data, MLL achieves fairly good performance both in LOS and NLOS area. Best result for RMSE of cluster power and location reduces to 3.49 dB and 2.06 m. The generalization experiments demonstrate that proposed model could be used to other scenarios without a significant drop in performance. 
Ablation study is also implemented to verify the important role of each module in proposed model.
Additionally, CIRs are regenerated utilizing the predicted cluster center and power, accurately matching MPC at different scales.
In the future, we will continue to study and analyze the channel characteristics super-resolution issue in depth, looking forward to finding better rules and strategies to achieve higher-quality and real-time CIR reconstruction for mobile channel modeling.
\balance

\section*{Acknowledgment}
This work is supported by National Key R$\&$D Program of China under Grant 2020YFB1806604, NSFC under Grant 62271043, the Ministry of Education of China under Grant 8091B032123, ZTE Corporation and the State Key Laboratory of Mobile Network and Mobile Multimedia Technology.
\bibliographystyle{IEEEtran}

\end{document}